\documentclass[fleqn,usenatbib]{mnras}

\usepackage{amsmath}
\usepackage{amssymb}	
\usepackage{mathptmx}
\usepackage{txfonts}

\usepackage[T1]{fontenc}

\DeclareRobustCommand{\VAN}[3]{#2}
\let\VANthebibliography\thebibliography
\def\thebibliography{\DeclareRobustCommand{\VAN}[3]{##3}\VANthebibliography}

\usepackage{graphicx}	

\usepackage{array}
\usepackage{varwidth}
\usepackage{makecell}
\newcolumntype{x}[1]{>{\centering\arraybackslash}p{#1}}

\usepackage{nicematrix}

\usepackage{tikz}

\newcommand{\z}[1]{\mbox {ZZ PsA}}

\def\p0{\phantom{0}}

\usepackage{xcolor}

\title[Metallicity and Contact Binary Mass Ratio]{Effects of Metallicity on the Instability Mass Ratio of Low Mass Contact Binary Systems.}

\author[S. S. Wadhwa et al.]{Surjit S. Wadhwa,$^{1}$\thanks{E-mail: 19899347@student.westernsydney.edu.au}
Natália R. Landin,$^{2}$
Petar Kostić,$^{4}$
Oliver Vince,$^{4}$
Bojan Arbutina,$^{3}$
\newauthor Ain Y. De Horta,$^{1}$
Miroslav D. Filipovi\'c,$^{1}$ Nicholas F.H. Tothill,$^{1}$
Jelena Petrovi\'c$^{4}$ and
\newauthor Gojko Djura\v sevi\'c$^{4}$
\\
$^{1}$School of Science, Western Sydney University, Locked Bag 1797, Penrith, NSW 2751, Australia\\
$^{2}$Universidade Federal de Viçosa, Campus UFV Florestal, CEP 35690-000 Florestal, MG, Brazil\\
$^{3}$Department of Astronomy, Faculty of Mathematics, University of Belgrade, Studentski trg 16, 11000 Belgrade, Serbia\\
$^{4}$Astronomical Observatory, Volgina 7, 11060 Belgrade, Serbia\\
}

\date{Accepted XXX. Received YYY; in original form ZZZ}

\pubyear{2023}

\begin{document}
\setlength{\extrarowheight}{0.05cm}
\label{firstpage}
\pagerange{\pageref{firstpage}--\pageref{lastpage}}
\maketitle

\begin{abstract}
The orbital stability of contact binary systems has been receiving considerable attention recently. Theoretical studies indicate that merger is likely to occur at very low mass ratios, but the actual mass ratio at which merger may take place is likely to be variable and dependent on the mass of the primary. We consider the effects of metal content on the orbital stability of contact
binary systems by modelling the gyration radius of a rotating and tidally distorted primary component at various values of $\rm [Fe/H]$ in the range -1.25 to +0.5. We determine the instability mass ratio range for contact binary systems with a low mass primary $0.6 \rm M_{\odot}\leq M_1 \leq 1.4\rm M_{\odot}$ at various metallicity levels and show that systems with low metallicity have an instability mass ratio lower than those with higher metal content and therefore are likely to be more stable. We illustrate the effect through light curve analysis of two otherwise very similar contact binary systems, except for different metallicity. While both would be considered unstable if metallicity was not taken into consideration, only one remains in that category after appropriate adjustments based on metallicity have been made.

\end{abstract}




\section{Introduction}
 \label{sec:intro}

Contact binary systems are common among close binaries with estimates suggesting that 1 in 500 stars in the galaxy disk are contact binaries \citep{2006MNRAS.368.1319R}. Their potential merger has gained significant interest since the recognition that transients such as luminous red novae are the result of the merger of components in contact binary systems \citep{2011A&A...528A.114T}. Although the galactic frequency of such mergers is thought to be as high as once every two to three years, the frequency of observable events is thought to be near once per decade \citep{2014MNRAS.443.1319K}. Nova Sco-2008 (=V1309 Sco) remains the only confirmed case of a contact binary merger event \citep{2011A&A...528A.114T}. Other examples such as V838 Mon \citep{2002IAUC.7785....1B}, OGLE2002-BLG-360 \citep{2013A&A...555A..16T} and V4432 Sgr \citep{1999AJ....118.1034M} are likely to represent stellar merger events although their progenitors remain unidentified. All recognised red novae have been observed post event. None, including V1309 Sco, had been recognised as potential merger candidates; so there were no detailed pre-merger observations. The lack of pre-merger data has intensified interest in the investigation of orbital stability of contact binary systems \citep{2021MNRAS.501..229W, 2021MNRAS.502.2879G, 2022MNRAS.512.1244C, 2023MNRAS.519.5760L} with the aim of identifying and studying systems prior to merger. \citet{1995ApJ...438..887R}, {\citet{2007ApJ...662..596L} and \citet{2007MNRAS.377.1635A, 2009MNRAS.394..501A}} have demonstrated that merger events will take place when the mass ratio of the components is quite low and \citet{2021MNRAS.501..229W} linked the instability mass ratio and separation to the mass of the primary ($M_1)$ component and concluded that the instability mass ratio can range from below 0.05 to above 0.2 for systems where $0.6\rm M_{\odot}\leq M_1\leq1.6\rm M_{\odot}$. 

Among the key parameters determining the instability mass ratio are the gyration radii of the components ($k_1, k_2$). 
Although it has been known for some time that the gyration radius of a star is dependent on its mass and composition \citep{1988AJ.....95.1895R, 2004A&A...424..919C, 2009A&A...494..209L, 2019A&A...628A..29C} the full impact of the gyration radius of the primary on orbital stability was only recently characterised \citep{2021MNRAS.501..229W}. The gyration radius of a star\, $k$, defined as $k=\sqrt{I/MR^2}$,  is proportional to its moment of inertia ($I$) and as such dependent on the mass distribution, opacity and energy generation rate of the star. The mass distribution as well as the opacity depend on the composition of the star particularly the presence of heavier elements or more generally on its metallicity. The effect of metallicity on the moment of inertia and gyration radius has not been well explored; \citet{2020ApJ...889..108A} suggest that for solar-like stars ($0.7 - 1.3\rm M_{\odot}$), metal poor stars have smaller gyration radius. Similarly, the effects of metallicity on orbital stability have also not received much attention. \citet{2010Ap&SS.329..283J} considered the effects of metallicity on the minimum mass ratio of a contact binary system with a $1.2\rm M_{\odot}$ primary. They confirmed that lower metallicity results in a lower minimum mass ratio, however, effects of using different values of the gyration radii of the primary and secondary components on orbital stability were not explored.

Most contact binaries are of low mass and mainly spectral class F, G and K. The brightest observable examples therefore are likely to be within the Solar neighborhood. \citet{1995PASP..107..648R} and \citep{2013AJ....146...70R} note that the metallicity range of Galactic disk contact binaries is expected to be $-0.5\leq\rm [Fe/H]\leq0.5$. We reviewed over 8000 contact binaries observed by The Large Sky Area Multi-Object Fiber Spectroscopic Telescope (LAMOST) as catalogued in \citet{2020RAA....20..163Q} and find that over 80\% have metallicity in the $-0.5\leq\rm [Fe/H]\leq0.5$ range. Most others have lower metallicity with very few having metallicity higher than 0.5. In this study, we explore the effects of metallicity (in the expected range) on the orbital stability parameters for contact binaries with primaries in the mass range $0.6\rm M_{\odot}\leq M_1\leq1.4\rm M_{\odot}$. We model the gyration radii of rotating and tidally distorted stars from $0.1\rm M_{\odot}$ to $1.4\rm M_{\odot}$ in increments of $0.02\rm M_{\odot}$ up to $0.2\rm M_{\odot}$ and thereafter in increments $0.1\rm M_{\odot}$ for metallicities $\rm [Fe/H] =$ -1.25 to 0.5 in increments of 0.125. We then insert the modelled gyration radii in the instability mass ratio equations from  \citet{2021MNRAS.501..229W} to determine the effects of metallicity on the instability mass ratio range for low mass contact binary systems. 

The paper is divided into 5 sections. Section 2 briefly outlines the ATON stellar evolution code used to calculate the gyration radii and discusses the results; section 3 describes the effect of metallicity on orbital stability parameters; section 4 explores the effect of metallicity-corrected gyration radii on orbital stability of two near-identical poorly studied contact binaries, and section 5 provides a brief summary and concluding remarks.

\section{Metallicity and the gyration radii}
 \label{sec:S2}
 Effects of metallicity on the internal structure of low mass stars, particularly incorporating the effects of rotation and tidal distortion, are poorly characterised. Recent developments suggest that structural detriments such as the gyration radius play a crucial role in the orbital stability of typical contact binary systems. We present, for the first time, modelling of the gyration radii of low mass stars incorporating tidal distortion and tidal instability at various internal compositions. The modelled data is used to explore the effects of internal composition on the orbital stability of the most common type of contact binary systems.
 \subsection{ATON - Stellar Evolution Code}
 Briefly, we use the {\ttfamily ATON} evolutionary code \citep{landin06,2009A&A...494..209L} to model the gyration radii of low mass stars from $0.1\rm M_{\odot}$ to $1.4\rm M_{\odot}$ including the effects of rotation and tidal distortion. In our models, convection is treated according to the traditional Mixing Length Theory \citep[][with the convection efficiency parameter $\alpha$=1.5]{vitense58}.
Grey atmosphere models are used to obtain surface boundary conditions that matches the interior solution at optical depth $\tau=2/3$. We used the opacities reported by \citet{rogers1} and \citet{1994ApJ...437..879A} and the equations of state from \citet{rogers2} and \citet{mihalas}. Here, we assume the Solar chemical composition X=0.7125 and Z=0.0175 \citep[taken from][under the error bars]{anders89} and that the elements are mixed instantaneously in convective regions. Our models were generated by considering rigid body rotation \citep{mendes99}. The initial angular momentum of each model was obtained according to the \citet{kawaler87} relation
%
\begin{equation}
J_{\rm Kaw}=1.566 \times 10^{50} \left( {M \over \rm M_{\odot}} \right) ^{0.985}
~~~{\rm g~cm^2~s^{-1}}. 
\end{equation}
%
For a more detailed description of the modelling code the reader is directed to \citet{2009A&A...494..209L}.

\subsection{Gyration radii of the secondary component}
The modelled gyration radii for stars from $0.1\rm M_{\odot}$ to $1.4\rm M_{\odot}$ are summarised in Table 1 and select samples plotted in Figure 1.

\begin{figure}
    \label{fig:F1}
	\includegraphics[width=\columnwidth]{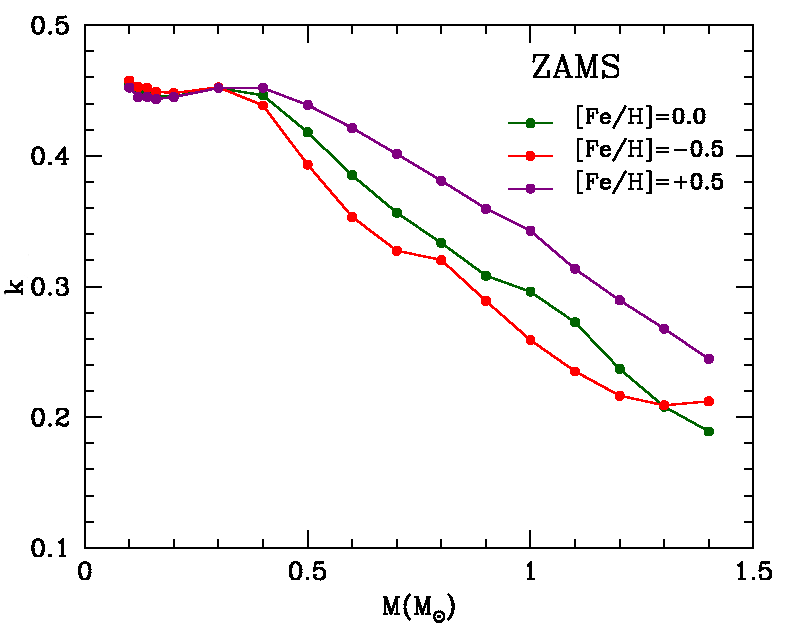}
    \caption{Gyration radius dependence on mass, for stars of in the mass range $0.1{\rm \rm M_{\odot}}$ to $1.4{\rm \rm M_{\odot}}$ at metallicities $[{\rm Fe/H}]=$ 0.5, 0 and -0.5.}
    \end{figure}

\begin{table*}
    \centering
   \scriptsize
   \advance\tabcolsep by -2pt

    \caption{Gyration radii of stars from $0.6{\rm \rm M_{\odot}}$ to $1.4{\rm \rm M_{\odot}}$ at metallicities from $[{\rm Fe/H}]$ -1.250 to 0.500 at the ZAMS.}
    \begin{NiceTabular}{|c|c|c|c|c|c|c|c|c|c|c|c|c|c|c|c|}
    \hline
        \diagbox{\hspace{-0.3ex}$\scriptscriptstyle {{{{M\mathrm{(\rm M_{\odot})}}}}}$}{$\scriptscriptstyle \mathrm{[Fe/H]}$}&
 -1.250 & -1.125 & -1.000 & -0.875 & -0.750 & -0.625 & $\vphantom{\big|}$ -0.500 & -0.375 & -0.250 & -0.125 & 0.000 & 0.125 & 0.250 & 0.375 & 0.500 \\ \hline
~~~~~~~0.10~~~~~~~~ & 0.458881 & 0.458740 & 0.458539 & 0.458354 & 0.458050  & 0.457929 
       & 0.457725 & 0.457442 & 0.457053 & 0.456729 & 0.456131 & 0.455536 & 0.454747 & 0.453731 & 0.452256 \\ 
         0.12 & 0.454348 & 0.454151 & 0.453847 & 0.453567 & 0.453151 & 0.453058 
       & 0.452816 & 0.452479 & 0.45206  & 0.451011 & 0.449788 & 0.449177 & 0.448389 & 0.446100 & 0.444892 \\ 
         0.14 & 0.453740 & 0.453517 & 0.453268 & 0.453013 & 0.452603 & 0.452519 
       & 0.452266 & 0.451960 & 0.451544 & 0.450143 & 0.448622 & 0.447963 & 0.447142 & 0.446238 & 0.445077 \\ 
         0.16 & 0.449953 & 0.449880 & 0.449664 & 0.449521 & 0.449287 & 0.449352 
       & 0.449029 & 0.448938 & 0.448708 & 0.446920 & 0.445408 & 0.444990 & 0.444557 & 0.444026 & 0.443451 \\ 
          0.20 & 0.448691 & 0.448670 & 0.448460 & 0.448429 & 0.448448 & 0.448223 
       & 0.448115 & 0.448022 & 0.447875 & 0.446723 & 0.445676 & 0.445534 & 0.445537 & 0.445243 & 0.445030 \\ 
          0.30 & 0.452659 & 0.452611 & 0.452586 & 0.452549 & 0.452503 & 0.452452 
       & 0.452411 & 0.452347 & 0.451628 & 0.452106 & 0.452045 & 0.451974 & 0.451922 & 0.451878 & 0.451824 \\ 
          0.40 & 0.436239 & 0.436259 & 0.435871 & 0.435958 & 0.436004 & 0.437263 
       & 0.438583 & 0.440382 & 0.441741 & 0.444180 & 0.446404 & 0.447837 & 0.449471 & 0.450853 & 0.452058 \\ 
          0.50 & 0.381676 & 0.382230 & 0.382526 & 0.383856 & 0.386152 & 0.389224 
       & 0.393337 & 0.398522 & 0.404450 & 0.411224 & 0.418111 & 0.423443 & 0.429188 & 0.433934 & 0.438948 \\ 
          0.60 & 0.348895 & 0.348608 & 0.347318 & 0.347514 & 0.347353 & 0.349814 
       & 0.353162 & 0.358790 & 0.366243 & 0.375455 & 0.385221 & 0.393649 & 0.403253 & 0.411774 & 0.421226 \\ 
          0.70 & 0.318230 & 0.318795 & 0.322101 & 0.319967 & 0.321381 & 0.323784 
       & 0.327393 & 0.332122 & 0.337926 & 0.346153 & 0.356586 & 0.366468 & 0.378031 & 0.388882 & 0.401472 \\ 
          0.80 & 0.302708 & 0.304529 & 0.306971 & 0.309561 & 0.312373 & 0.316234 
       & 0.320210 & 0.324785 & 0.315970 & 0.322925 & 0.333341 & 0.342850 & 0.354542 & 0.366085 & 0.380913 \\ 
          0.90 & 0.265800 & 0.267263 & 0.269609 & 0.273330 & 0.278835 & 0.283057 
       & 0.289037 & 0.296096 & 0.304744 & 0.313319 & 0.308374 & 0.319583 & 0.332718 & 0.344908 & 0.359595 \\ 
          1.00 & 0.240938 & 0.241431 & 0.242413 & 0.244797 & 0.249186 & 0.253175 
       & 0.259141 & 0.266342 & 0.277531 & 0.285166 & 0.296167 & 0.305630 & 0.317291 & 0.328507 & 0.342763 \\ 
          1.10 & 0.228613 & 0.226599 & 0.224680 & 0.224593 & 0.226246 & 0.229759 
       & 0.235026 & 0.245127 & 0.255054 & 0.261301 & 0.272595 & 0.279257 & 0.290725 & 0.299426 & 0.313488 \\ 
          1.20 & 0.230766 & 0.227028 & 0.222618 & 0.217458 & 0.214213 & 0.211588 
       & 0.216479 & 0.219495 & 0.226778 & 0.230794 & 0.237031 & 0.245471 & 0.257358 & 0.270364 & 0.289545 \\ 
          1.30 & 0.234175 & 0.230621 & 0.228894 & 0.220633 & 0.214296 & 0.209778 
       & 0.209098 & 0.205601 & 0.208048 & 0.198791 & 0.208123 & 0.216998 & 0.231171 & 0.245683 & 0.267786 \\ 
          1.40 & 0.237448 & 0.233850 & 0.237964 & 0.224308 & 0.224154 & 0.213303 
       & 0.212105 & 0.207452 & 0.183293 & 0.182651 & 0.189123 & 0.197858 & 0.210357 & 0.224507 & 0.244657 \\ \hline
        
    \end{NiceTabular}
   
\end{table*}

Orbital instability is thought to occur at very low mass ratios. Even for the smallest stars, where the instability ratio may be as high as 0.22, the mass of the secondary is well below $0.2\rm M_{\odot}$. Below a certain mass a star becomes fully convective as the inner radiative core disappears. As shown by \citet{2000ARA&A..38..337C} this mass is found to be $0.35\rm M_{\odot}$ for metallicities $\rm [Fe/H] \geq-2.0$. The fully convective mass limit does drop in the case of very metal poor stars reaching as low as $0.286\rm M_{\odot}$ for $\rm [Fe/H] \approx -3.0$ \citep{2021A&A...650A.184M}. It is clear however that, regardless of the metallicities of contact binaries likely to be encountered, the secondary can be classified as being fully convective. Being fully convective the gyration radius is not expected to change significantly and this is reflected in the modelled values for low mass stars $\leq0.2\rm M_{\odot}$ (see Table 1) for which the mean value for the entire range of modelled metallicities is $0.4494\pm0.004$. The value is in good agreement with the fully convective polytrope ($n=1.5$) value of 0.4527 for the gyration radius. For the remainder of this study we have chosen to adopt the fully convective polytrope value for the gyration radius of the secondary.

\subsection{Gyration radius of the primary component}
Previous estimates for gyration radii of non rotating stars of $M\geq 0.8~{\rm M_{\odot}}$ at different matallicities suggest a general trend of reduction in the gyration radius with increasing mass at any given metallicity and reduction in gyration radius with reducing metallicity for any given mass \citep{2019A&A...628A..29C}. A similar trend is obvious (see Table 1 and Figure 1) in our modelling of rotating and tidal-distorted low mass stars from $0.6\rm M_{\odot}$ to $1.4\rm M_{\odot}$ for metallicities $-0.25\leq \rm [Fe/H]\leq0.5$ at the Zero Age Main Sequence (ZAMS). In fact, very good linear fits can be obtained at all metallicities in this range:
{
\begin{equation}
    k_1 = a{\rm M_1} + b.
\end{equation}
}
\noindent We provide the coefficients $a,b$ and goodness of fit parameter ($R^2$) in Table 2. The situation is somewhat more complex at lower metallicities. As metallicity drops the gyration radius actually increases for higher mass stars such that the linear correlation only holds up to $M=1.2\rm M_{\odot}$ for metallicities -0.375 and -0.5 (coefficients included in Table 2) with the trend continuing such that by $\rm [Fe/H] = -1.0$ the linear correlation only holds to $M=1.1\rm M_{\odot}$. 
\begin{table}
    \caption{Linear fit coefficients and goodness of fit parameter for the gyration radius-mass for metallicities between -0.5 to 0.5. The linear relation hold for stars with masses $0.6\rm M_{\odot}$ to $1.4\rm M_{\odot}$ except for $\rm [Fe/H] = -0.375$ and $-0.5$ where the maximum stellar mass is $1.2\rm M_{\odot}$}.
    \centering
    \begin{tabular}{|c|c|c|c|}
    \hline
        \rm [Fe/H] & $a$ & $b$ & $R^2$ \\ \hline
        0.500 & -0.2119 & 0.5464 & 0.995 \\ 
        0.375 & -0.2296 & 0.5489 & 0.997 \\ 
        0.250 & -0.2397 & 0.5480 & 0.997 \\ 
        0.125 & -0.2484 & 0.5452 & 0.997 \\ 
        0.000 & -0.2524 & 0.5409 & 0.995 \\ 
        -0.125 & 0.2544 & 0.5355 & 0.993 \\ 
        -0.250 & -0.2455 & 0.5229 & 0.989 \\ 
        -0.375 & -0.2608 & 0.5288 & 0.983 \\ 
        -0.500 & -0.2650 & 0.5267 & 0.981 \\\hline
    \end{tabular}
\end{table}

If we look at the gyration radius-mass dependence for $\rm [Fe/H] = -1.25$ (green profile in Figure 2), we note a definite upswing at $M\approx1.1\rm M_{\odot}$. The orange profile represents the gyration radius of rotating and tidally distorted stars from $0.9\rm M_{\odot} < \emph{M} <2.3\rm M_{\odot}$ at Solar metallicity as originally published by \citet{2009A&A...494..209L}. It is clear that a similar upswing is present but at a higher mass level of $M\approx 1.5\rm M_{\odot}$. It was postulated by \citep{1989A&AS...81...37C} and \citep{2009A&A...494..209L} that the upswing in the gyration radius was due to a change in the internal structure of the star due to the ignition and transition to the Carbon-Nitrogen-Oxygen (CNO) burning cycle as opposed to the Proton-Proton (P-P) chain normally seen in lower mass stars. 

The main factor determining the transition to CNO over P-P is the central temperature. Typically, $logT_c \approx 7.22$ is considered the temperature where the CNO process is transitioning to the dominant energy source \citep{2009ApJ...701..837S}. We modelled the central temperature of stars from $0.9\rm M_{\odot} \leq \emph{M} \leq 1.4\rm M_{\odot}$ at different metallicities. Figure 3 illustrates the change in central temperature at the ZAMS with change in metallicity. It is clear that the central temperature is higher in all cases in metal poor stars. Heavier stars reach CNO transition temperatures just below Solar metallicity while a $1.1\rm M_{\odot}$ star reaches CNO transition temperature at $\rm [Fe/H] = -1.25$, confirming that CNO transition is likely triggered at lower masses in metal poor stars and potentially explains the change in gyration radius profile at lower metallicities. As noted above, over 80\% of contact binary systems with spectroscopically determined metallicity lie within the $-0.5 \leq \rm [Fe/H] \leq 0.5$ range such that the linear coefficients in Table 2 should be sufficient.

\begin{figure}
    \label{fig:F2}
	\includegraphics[width=\columnwidth]{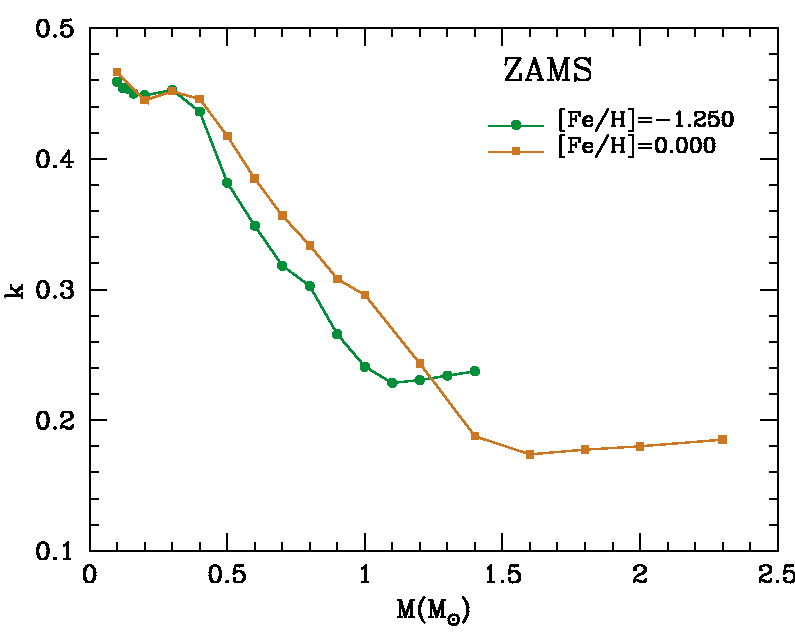}
    \caption{Gyration radius-mass dependence for stars with mass ranging from $0.1\rm M_{\odot}$ to $1.4\rm M_{\odot}$ at metallicity $\rm [Fe/H]=$ -1.25 and for stars $0.1\rm M_{\odot}$ to $2.3\rm M_{\odot}$ at metallicity $\rm [Fe/H]=$ 0 from \citet{2009A&A...494..209L} at the ZAMS.}
    \end{figure}

    \begin{figure}
    \label{fig:F3}
	\includegraphics[width=\columnwidth]{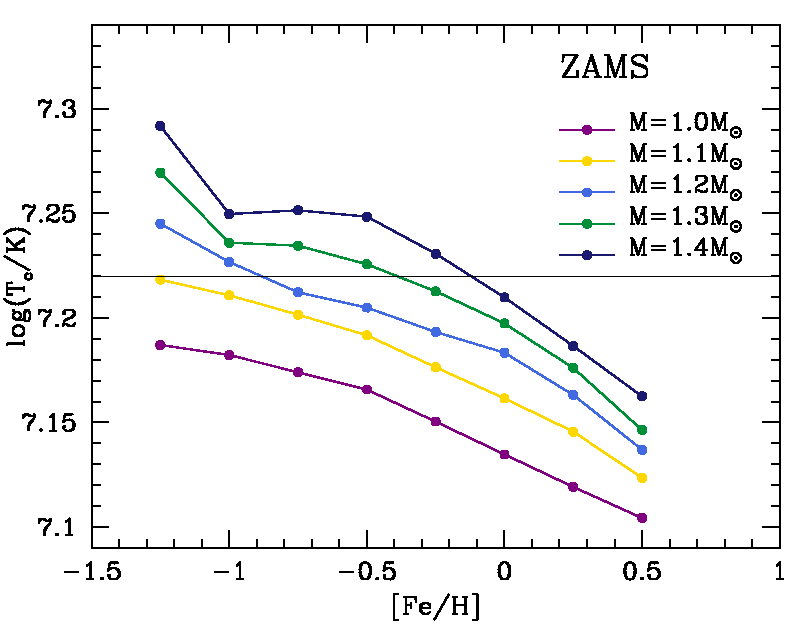}
    \caption{Central temperature of stars from $1.0\rm M_{\odot}$ to $1.4\rm M_{\odot}$ at the ZAMS for various metallicities.}
    \end{figure}

 \section{Instability Mass Ratio and Metallicity}

 In addition to rising central temperature, reduced metallicity also leads to a reduction in stellar radius especially for stars more massive than $1.1\rm M_{\odot}$ \citep{2019A&A...628A..29C}. The Roche geometry model of contact binary systems requires both components to fill their respective Roche lobes such that they are surrounded by a common envelope of essentially equal temperature. For a low metallicity systems to be in a contact configuration the components must be closer together and have smaller periods. The instability parameters are strongly influenced by the distribution of mass within a star (i.e gyration radius) and as noted above the gyration radius changes significantly with metallicity, so we expect a significant change in orbital stability parameters at different metallicity levels. In this section we explore the effects of metal content on the instability mass ratio of contact binary systems with mass of the primary ranging from $0.6\rm M_{\odot}$ to $1.4\rm M_{\odot}$. 
 
  The instability mass ratio ($q_{\rm inst}$) can be found numerically from equation \citep{2007MNRAS.377.1635A, 2021MNRAS.501..229W}.



\begin{equation}
\begin{array}{@{\extracolsep{-3mm}} lll @{}}
\frac{q\frac{k_2^2}{k_1^2}{P}{Q} + \sqrt{(q\frac{k_2^2}{k_1^2}{P}{Q})^2 + 3 (1+q\frac{k_2^2}{k_1^2}{Q}^2) (q \frac{k_2^2}{k_1^2}{P}^2 + \frac{q}{(1+q)k_1^2})   }} {q \frac{k_2^2}{k_1^2}{P}^2 + \frac{q}{(1+q)k_1^2}} =
  \frac{0.6q^{-2/3} + \ln (1+ q^{-1/3})}{0.49q^{-2/3}+0.15 f},
 \end{array}
\end{equation}

\noindent where $k_{1,2}$ are the gyration radii of the components, $f$ is the fill-out factor and 

\begin{footnotesize}

\begin{equation}
P = \frac{0.49q^{2/3}-3.26667q^{-2/3}(0.27q -0.12q^{4/3})}{0.6q^{2/3} + \ln (1+ q^{1/3})} 
\end{equation}
\end{footnotesize}
and
\begin{footnotesize}
\begin{equation}
Q = \frac{(0.27q -0.12q^{4/3})({0.6q^{-2/3} + \ln (1+ q^{-1/3})})}{0.15 (0.6q^{2/3} + \ln (1+ q^{1/3}))}.
\end{equation}
\end{footnotesize}\\

\noindent One can obtain the instability mass ratio range for any system by solving Equation 3 with $f$ fixed at 0 (marginal contact) or 1 (full overcontact).\\
 
 We solve Equation 3 for $q_{\rm inst}$ at $f_{0,1}$, using various values for $k_1$, for various metallicities from $-0.5 \leq \rm [Fe/H] \leq 0.5$ in steps $0.125$ as summarised in Table 3. We note a similar quadratic trend to that of \citet{2021MNRAS.501..229W} in the instability mass ratio for both fill-out values with higher instability mass ratio for lower mass primaries:
 
 \begin{equation}
     q_{\rm inst} = c{\rm M_1^2} + d{\rm M_1} + e.
 \end{equation}

\noindent We provide the quadratic fit coefficients and goodness of fit coefficient in Table 4 and representative graphics in Figure 4. The quadratic fits allow determination of the instability mass ratio range without having to solve Equation 3. The progression from $f=0$ to $f=1$ is linear in all cases so it is easy to estimate the instability mass ratio at any given value of $f$.
 
 Reviewing the summary table for instability mass ratios, it is clear that metal-poor systems have a significantly lower instability mass ratio relative to Solar metallicity systems. For example, for a system with a Solar mass primary the instability mass ratio range at $\rm [Fe/H] = -0.5$ is 0.085 - 0.098 rising to 0.138 - 0.168 at $\rm [Fe/H] = 0.5$. Current reports of potential red nova progenitors which have considered differing values for the gyration radii have relied on Solar metallicity estimates. The actual physics is clearly more complex and effects of metal content needs due consideration when investigating the orbital stability of contact binary systems.

 \begin{table*}
    \centering
    \scriptsize
    \caption{Instability mass ratio fit parameters for stellar masses $0.6{\rm \rm M_{\odot}}$ to $1.4{\rm \rm M_{\odot}}$ and metallicities from $[{\rm Fe/H}]$ -0.5 to 0.5.}
        \begin{NiceTabular}{|c|c|c|c|c|c|c|c|c|c|}
    \hline
    \diagbox{\hspace{-0.3ex}$\scriptscriptstyle {{{{\mathrm{[Fe/H]}{}}}}}$}{$\scriptscriptstyle \mathrm{(\rm M_{\odot})}$}&
         0.6  & 0.7 & 0.8 & 0.9 & $\vphantom{\big|}$ 1.0 & 1.1 & 1.2 & 1.3 & 1.4  \\ \hline
       ~~~ -0.500 ~~~ & 0.146 - 0.179 & 0.128 - 0.154 & 0.123 - 0.147 & 0.103 - 0.121 & 0.085 - 0.098 & 0.072 - 0.081 & 0.062 - 0.070 & 0.059 - 0.065 & 0.060 - 0.067 \\ 
        -0.375 & 0.150 - 0.184 & 0.131 - 0.158 & 0.126 - 0.151  & 0.107 - 0.126 & 0.089 - 0.103 & 0.077 - 0.088 & 0.064 - 0.072 & 0.057 - 0.063 & 0.058 - 0.065  \\ 
        -0.250 & 0.155 - 0.192 & 0.135 - 0.164 & 0.127 - 0.153 & 0.113 - 0.133 & 0.096 - 0.111 & 0.083 - 0.095 & 0.068 - 0.076 & 0.058 - 0.065 & 0.047 - 0.051  \\ 
        -0.125 & 0.162 - 0.203 & 0.141 - 0.172 & 0.125 - 0.150 & 0.118 - 0.141 & 0.101 - 0.118 & 0.086 - 0.100 & 0.070 - 0.079 & 0.054 - 0.060 & 0.046 - 0.051  \\ 
        0.000 & 0.170 - 0.213 & 0.148 - 0.183 & 0.132 - 0.159 & 0.115 - 0.137 & 0.107 - 0.126 & 0.093 - 0.108 & 0.073 - 0.083 & 0.058 - 0.065 & 0.049 - 0.054  \\ 
        0.125 & 0.176 - 0.224 & 0.156 - 0.193 & 0.138 - 0.168 & 0.123 - 0.147 & 0.113 - 0.134 & 0.097 - 0.113 & 0.078 - 0.088 & 0.063 - 0.070 & 0.053 - 0.059 \\ 
        0.250 & 0.184 - 0.236 & 0.164 - 0.205 & 0.147 - 0.180 & 0.131 - 0.159 & 0.121 - 0.144 & 0.104 - 0.122 & 0.084 - 0.097 & 0.070 - 0.079 & 0.059 - 0.066  \\ 
        0.375 & 0.191 - 0.246 & 0.172 - 0.218 & 0.155 - 0.193 & 0.140 - 0.170 & 0.128 - 0.155 & 0.109 - 0.129 & 0.092 - 0.106 & 0.078 - 0.088 & 0.066 - 0.075  \\ 
        0.500 & 0.198 - 0.258 & 0.182 - 0.233 & 0.166 - 0.209 & 0.150 - 0.185 & 0.138 - 0.168 & 0.119 - 0.141 & 0.103 - 0.121 & 0.090 - 0.104 & 0.077 - 0.088 \\ \hline
    \end{NiceTabular}
\end{table*}

\begin{table}
    \centering
    \scriptsize
    \caption{Quadratic fit coefficients and goodness of fit parameter for the instability mass ratio-mass relation ($f=0,1$) for metallicities $\rm [Fe/H]$ 0.5 to -0.5. Note the quadratic fit parameters are from $M = 0.6\rm M_{\odot}$ to $1.4\rm M_{\odot}$ for $\rm [Fe/H]\geq$ -0.25 and up to $1.2\rm M_{\odot}$ for $\rm [Fe/H]<$ -0.25.}
    \begin{tabular}{|c|c|c|c|c|}
    \hline
        \rm [Fe/H] & $c$ & $d$ & $e$ & $R^2$ \\ \hline
        0.500 & 0.0129, 0.0547 & -0.1786, -0.3238 & 0.3007, 0.4330 & 0.999, 0.999 \\
        0.375 & 0.0192, 0.0601 & -0.1948, -0.3350 & 0.3000, 0.4242& 0.998, 0.998 \\ 
        0.250 & 0.0194, 0.0618 & -0.1946, -0.3338 & 0.2921, 0.4110 & 0.997, 0.997 \\ 
        0.125 & 0.0235, 0.0668 & -0.1998, -0.3374 & 0.2855, 0.3990 & 0.997, 0.996 \\ 
        0.000 & 0.0279, 0.0640 & -0.2048, -0.3231 & 0.2855, 0.9947 & 0.994, 0.995 \\ 
        -0.125 & 0.0153, 0.0484 & -0.1750, -0.2846 & 0.2591, 0.3523 & 0.995, 0.994 \\ 
        -0.250 & 0.0272, 0.0605 & -0.1872, -0.2932 & 0.2554, 0.3431 & 0.996, 0.996 \\ 
        -0.375 & -0.0036, 0.0167 & -0.1375, -0.2171 & 0.2914, 0.3073 & 0.991, 0.991 \\ 
        -0.500 & 0.0048, 0.0310 & -0.1521, -0.2421 & 0.2356, 0.3130& 0.988, 0.988 \\ \hline
    \end{tabular}
    
\end{table}

 \begin{figure}
    \label{fig:F4}
	\includegraphics[width=\columnwidth]{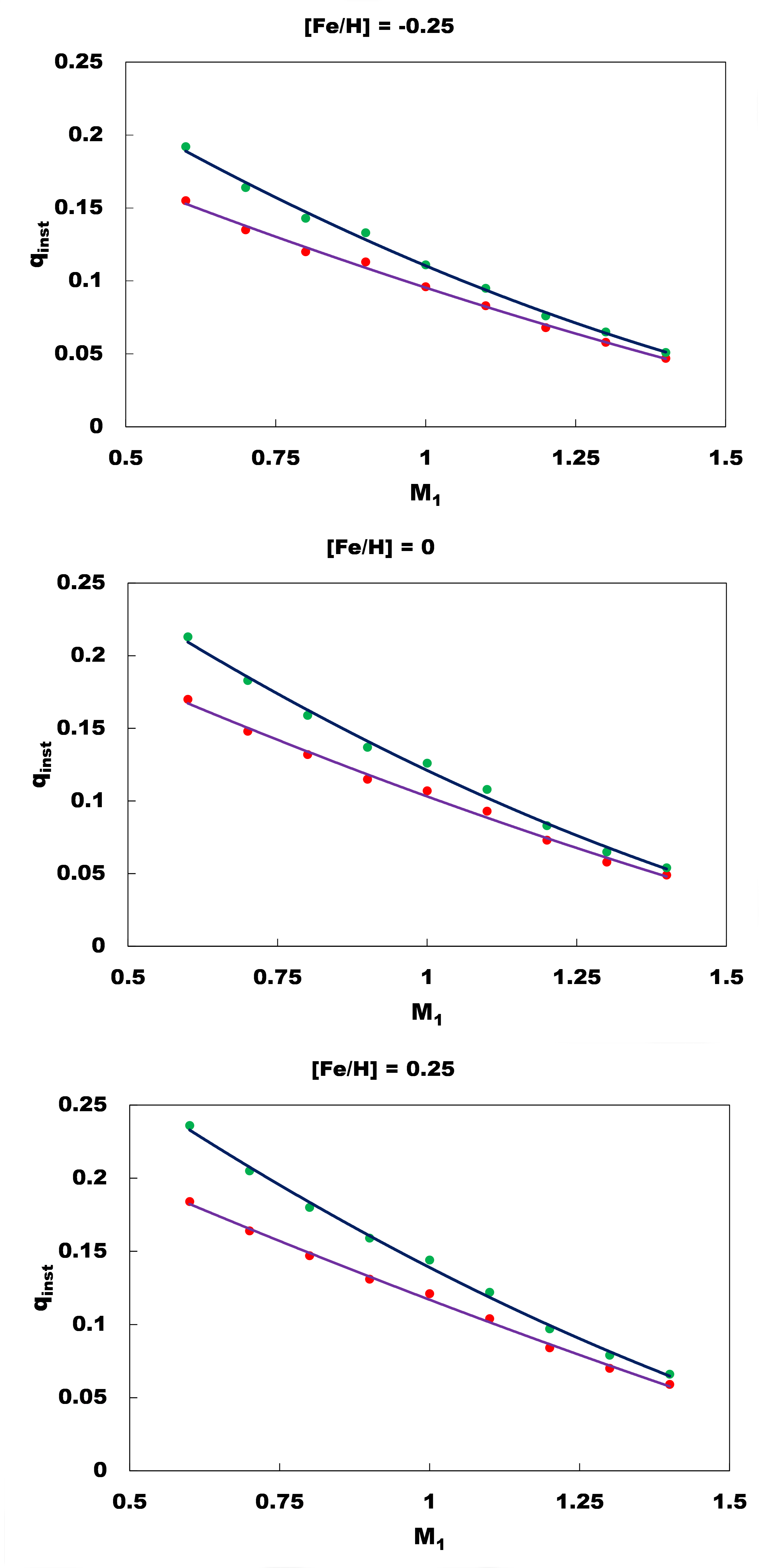}
    \caption{Quadratic fits at $f=$ 0 (red dots / purple line) and $f=$ 1 (green dots / blue line) at various metallicities. $ \rm M_1$ is Solar units. The quadratic fit cofficients are summarised in Table 4.}
    \end{figure}

 \section{Metallicity and Orbital stability - Practical Examples}
 \subsection{Observations}
 In this section we explore the effect of metallicity on the orbital stability of two very similar but poorly studied contact binaries. Both systems were selected from the All Sky Automated Survey - Super Nova (ASAS-SN) \citep{2014ApJ...788...48S, 2020MNRAS.491...13J}. The systems were selected based on techniques described in \cite{2022JApA...43...94W} which indicated that both were likely to be of extreme low mass ratio and potentially unstable.
 
 WISE J190843.4+373842 (hereafter W1908), ($\alpha_{2000.0} = 19\ 08\ 23.45$, $\delta_{2000.0} = +37\ 38\ 43.0$), also known as ASASSN-V J190843.45+373842.8, ZTF J190843.44+373842.8, was discovered by the Wide-field Infrared Survey Explorer (WISE) survey \citep{2018ApJS..237...28C} as a contact binary with the VSX database \citep{2006SASS...25...47W} reporting a period of 0.3092274 days. ASASSN-V J204400.26+575216.7 (hereafter A2044), ($\alpha_{2000.0} = 20\ 44\ 00.26$, $\delta_{2000.0} = +57\ 52\ 16.7$), also known as USNO-B1.0 1478-0400187, was discovered by the ASAS-SN with VSX database recording a period of 0.342893 days.
 
 W1908 was observed over 4 nights between July and August 2021 at the Vidojevica Observatory with the 140cm Telescope Milanković equipped with Andor iKon-L 936 CCD camera with a resolution of 0.39 arcs/pixel and standard Johnson B,V and R filters. In total, approximately 200 images were acquired in each pass-band. A2044 was observed over 3 nights in October 2022 at the Vidojevica observatory with the 60cm Telescope Nedeljković equipped with FLI ProLine PL23042 CCD camera with a resolution of 0.516 arcs/pixel and standard Johnson B,V and R filters. In total approximately 130 images were acquired in each pass-band.

 Aperture photometry was performed using the AstroImageJ software package \citep{2017AJ....153...77C} for each pass-band using 2MASS 1908509+3738284 and 2MASS 190884538+3737263 as the comparison and check stars respectively for W1908 and for A2044, 2MASS 20435919+5751541 and TYC 3959-43-1 respectively. For W1908 we find the V band amplitude variation 14.24 - 14.55 magnitude with the secondary eclipse at 14.52 magnitude. For A2044 the V band amplitude variation 11.5 to 11.8 magnitude with the secondary eclipse at 11.77 magnitude. Both systems were observed with the Transiting Exoplanet Survey Satellite (TESS). The TESS light curve for A2044 has an amplitude of 0.29 mag, similar to our V band observations. The TESS light curve for W1908 has considerable scatter which we believe is likely due to a combination of blending from a nearby fainter star approximately 7" distant, noting the TESS instrument has a pixel resolution of 21". In addition, W1908 is quite faint even when blended with the companion it is near the limit of the instrument.

 Based on survey V band data and our observations we refine the orbital ephemeris of W1908 as follows: 
\begin{center}
    
    $HJD_{ \rm min\ I} = 2459439.433007(313) + 0.3092266(50)E$\\                   

\end{center}
\noindent and of A2044 
\begin{center}
    
    $HJD_{ \rm min\ I} = 2459884.362481(52) + 0.3428950(40)E$.\\                   

\end{center}

 \subsection{Light Curve Analysis}
 Due to significant correlation between geometric parameters such as the mass ratio ($q$), inclination ($i$) and fill-out ($f$) the analysis of contact binary light curves to reliably determine geometric parameters, in the absence of radial velocity measurements, can only be carried out in the presence of total eclipses \citep{2005Ap&SS.296..221T}. Both of the systems under consideration demonstrate total eclipses and as such are suitable for light curve analysis. We use the 2013 version of the Wilson-Devenney light curve analysis code which incorporates Kurucz atmospheric models \citep{1990ApJ...356..613W, 1998ApJ...508..308K, 2021NewA...8601565N} to perform simultaneous analysis of B, V and R bands. The metallicity for the reported solution for W1908 was set to -0.7 (see below) during the analysis. We also analysed the TESS light curve for A2044 separately as the filter used by the TESS instrument is quite broad centered on the $I_c$ band  \citep{2015JATIS...1a4003R}. We modelled the TESS photometry using infrared limb darkening coefficients. As noted above the TESS light curve for W1908 was not suitable for accurate analysis. To overcome the strong correlation between the mass ratio and inclination, the mass ratio grid search method is employed where the light curve is analysed for various fixed values of $q$ and the search is refined to smaller and smaller intervals to find the correct mass ratio value. In our case we searched from $0.05 < q < 1.0$ in intervals of 0.1 and then refined the search to intervals of 0.01 around the first rough estimate and further still in increments of 0.001 near the second estimate. The analysis was performed with the temperature of the primary ($T_1$) fixed, see below, and the temperature of the secondary ($T_2$), inclination ($i$), dimensionless potential (fill-out), and the dimensionless luminosity of the primary ($L_1$) acting as the variable parameters. Iterations are carried out until the suggested corrections are less than the reported standard deviation. During the last iteration the mass ratio is also made a variable parameter and the standard deviation, as reported by the WD model, for each parameter was adopted as the potential error. Given the estimated temperature of the primary (see below) is less than $7200\rm K$, the gravity darkening coefficients were set as $g_1 = g_2 = 0.32$, the bolometric albedoes were set to $A_1 = A_2 = 0.5$, simple reflection treatment is applied and we used the 2019 updates of limb darkening coefficients from \citet{1993AJ....106.2096V}. The light curve solutions are summarised in Table 5. Figures 5 and 6 illustrate the observed and modelled light curves and the respective residuals from our observations in $B,V$ and $R$ bands respectively.

 \begin{figure}
    \label{fig:F5}
	\includegraphics[width=\columnwidth]{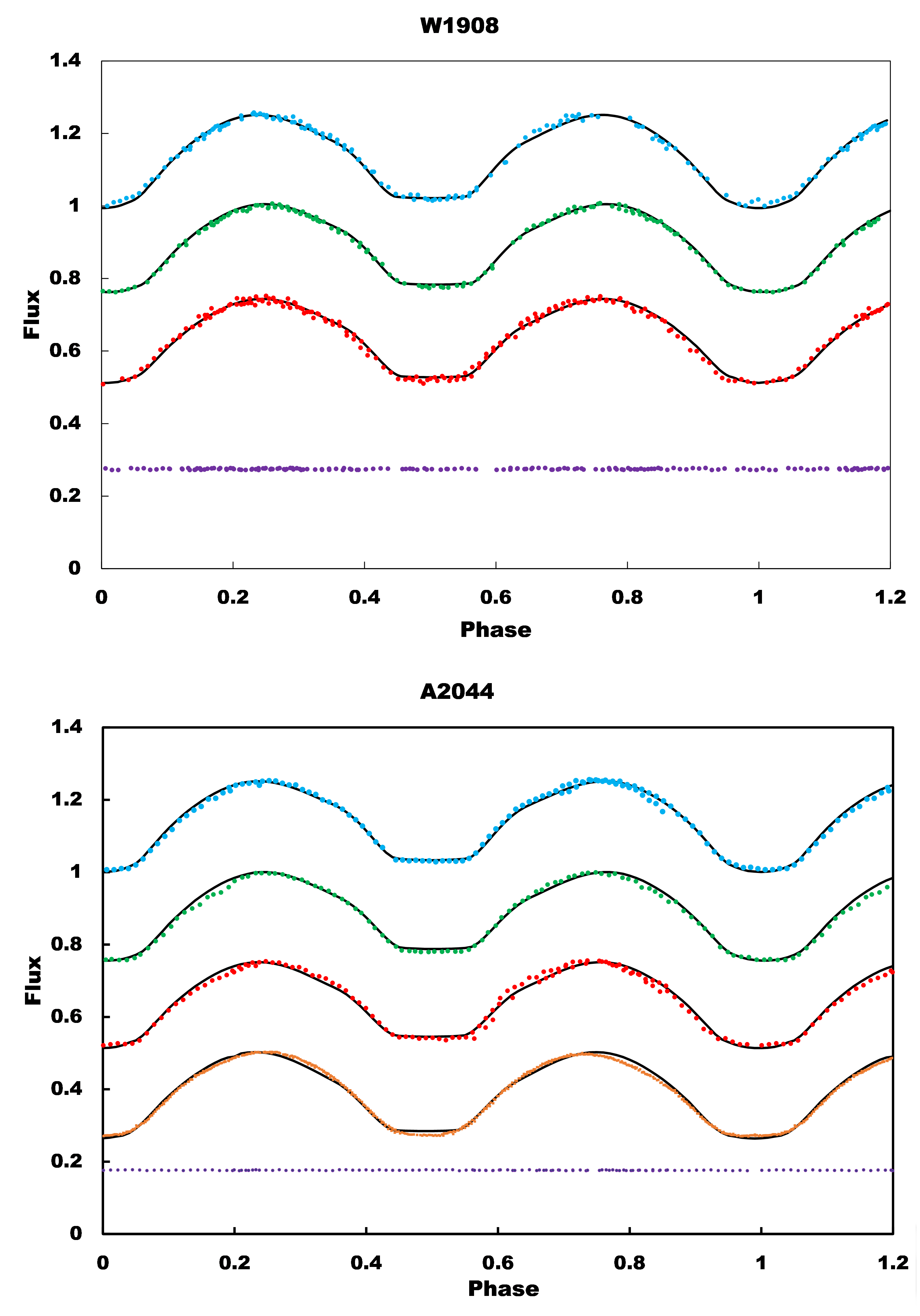}
    \caption{Observed and fitted light curves for W1908 and A2044. The black lines represent the WD fitted model while the blue, green and red dots represents the observed curves in various bands. The fitted orange curve for A2044 represents TESS observations. The purple dots represent the check star flux. The normalised flux has been shifted vertically for blue, red and TESS bands and the check star for clarity.}
    \end{figure}

     \begin{figure}
    \label{fig:F6}
	\includegraphics[width=\columnwidth]{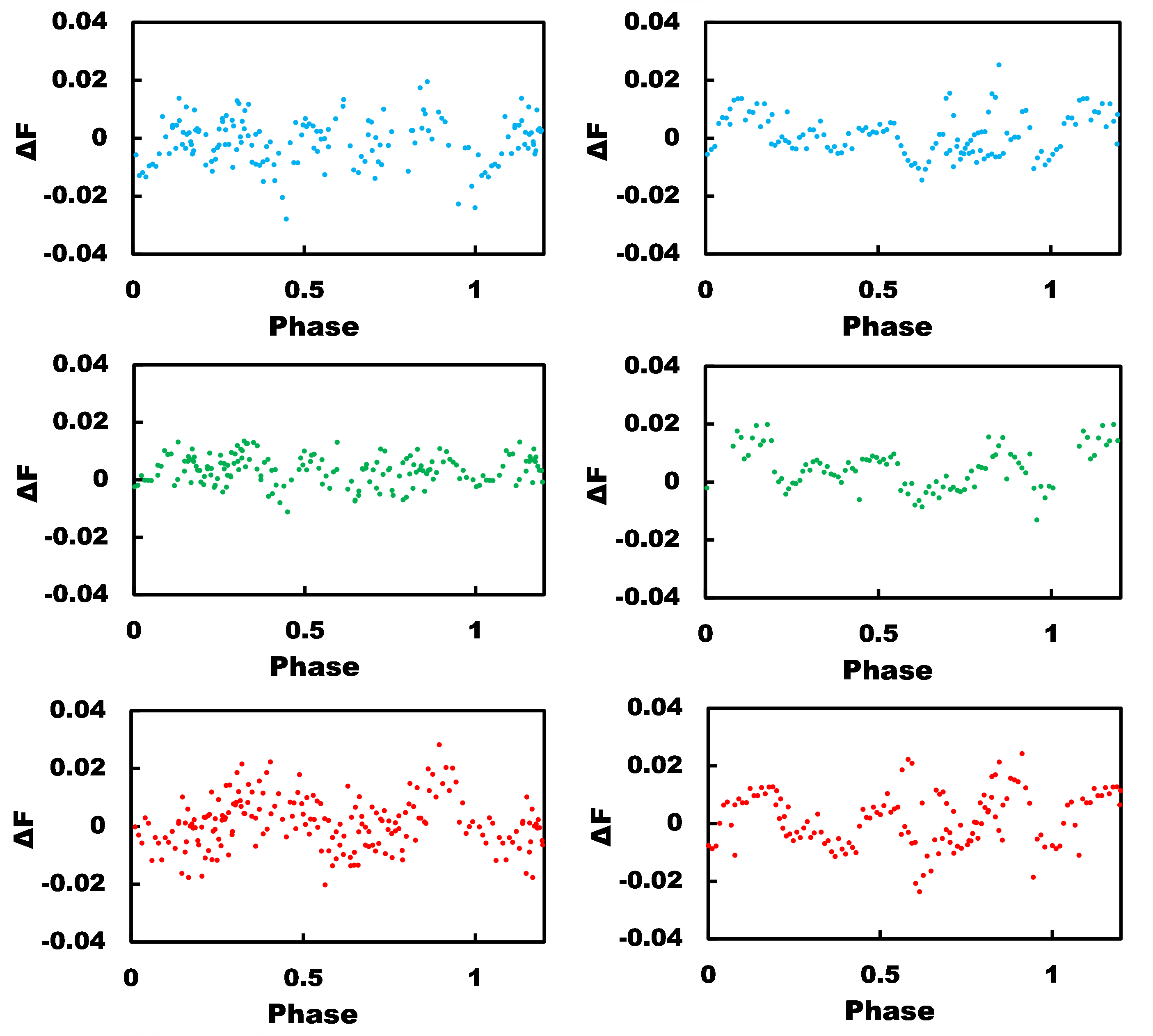}
    \caption{Residuals for W1908 (left) and A2044 (right) for ground based observations in $B,V$ and $R$ bands (top to bottom). Vertical scale ($\Delta$ F) represents the difference in normalised flux.}
    \end{figure}

 The presence of total eclipses places strong constraints on the geometric parameters and as such absolute temperature values of the primary and secondary have little influence on the light curve solution \citep{1993PASP..105.1433R, 2001AJ....122.1007R}. It is common practice to fix the temperature of the primary component during light curve analysis. Determining the fixed value, however, has been somewhat troublesome. Where available spectral class estimation is now preferred as photometric estimations can vary considerably \citep[see e.g.][]{2023PASP..135e4201L, 2023PASP..135d4201G, 2022MNRAS.517.1928G}. In our case the reported effective temperatures for W1908 on the VizieR database range from $5517\rm K$ to $6405\rm K$ and for A2044 from $5758\rm K$ to $6500\rm K$. \citet{2007ApJS..169..328R} and \citet{2008A&A...492..277B} compared the collective Spectral Energy Distribution (SED) from various photometric bands with synthetic theoretical spectra and found good agreement between spectral and SED estimated effective temperatures. We constructed SEDs for both systems from publicly available photometric data. The SEDs were fitted to theoretical spectra using the VO Sed Analyzer (VOSA) available through the Spanish Virtual Observatory \footnote{http://svo2.cab.inta-csic.es/theory/vosa/}. The SED estimated effective temperature of the primary for both systems was $6000\rm K$ which is not unexpected given the mass of the primary components (see below) are very similar. We adopted $6000\rm K$ as the fixed value for $T_1$ for both systems. The observed and fitted SED curves are illustrated in Figure 7.

  \begin{figure}
    \label{fig:F7}
	\includegraphics[width=\columnwidth]{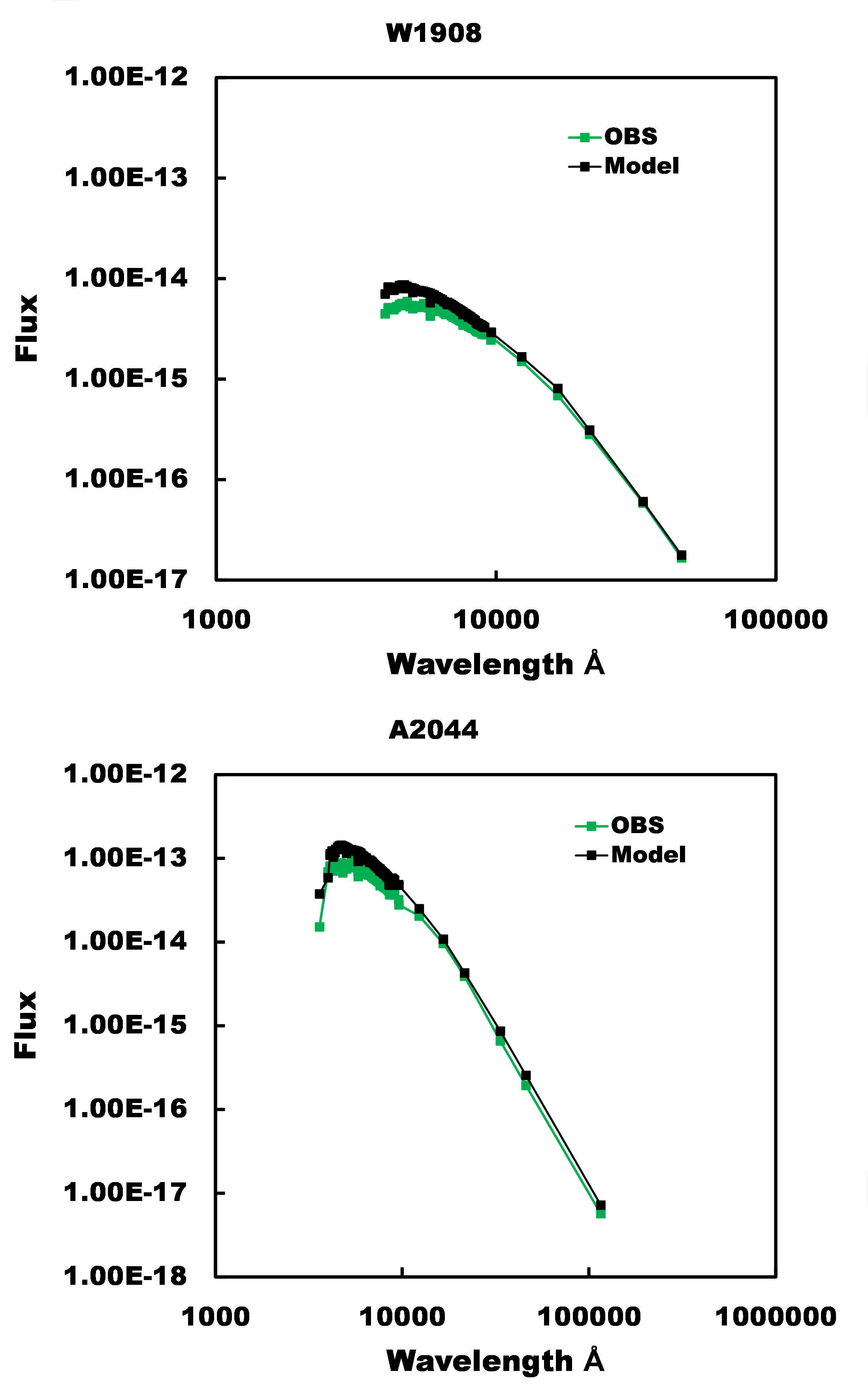}
    \caption{Observed and fitted SED for W1908 and A2044. The vertical axis flux in erg/cm$^2$/s/\AA. }
    \end{figure}

  \begin{table}
 
   \centering
\caption{Light curve solution and absolute parameters for W1908 and A2044. Main geometric parameter solution for the TESS light curve for A2044.}
   \begin{tabular}{ |l|l|l|l| }
    \hline
        \hfil  &\hfil W1908 &\hfil A2044 & \hfil A2044 (TESS) \\
        \hline
        \hfil $T_1$(K) (Fixed) &\hfil 6000 &\hfil 6000 &\hfil 6000 \\
        
        \hfil$T_2$(K) &\hfil $6006\pm19$ &\hfil $5893\pm21$ &\hfil $5846\pm16$ \\
         \hfil Incl. $(^\circ)$   &\hfil $79.8\pm0.9$  &\hfil $78.9\pm1.25$ &\hfil $77.6\pm0.55$\\ 
          \hfil $q$ &\hfil $0.111\pm0.001$  &\hfil $0.107\pm0.002$ &\hfil $0.106\pm0.002$ \\ 
           \hfil Fill out (\%) &\hfil $43\pm3$ &\hfil $49\pm3$ &\hfil $69\pm6$ \\
           \hfil $r_1$ (mean)  & \hfil 0.585 &\hfil 0.590 & \hfil0.598\\
           \hfil $r_2$ (mean)  & \hfil 0.227 &\hfil 0.227& \hfil 0.238\\
           \hfil $ M_1/\rm M_{\odot}$&\hfil $1.06\pm0.02$ &\hfil $1.04\pm0.02$&\\
           \hfil $ M_2/\rm M_{\odot}$&\hfil $0.12\pm0.02$ &\hfil $0.11\pm0.02$&\\ 
           \hfil $ M_{V1}$&\hfil $4.83\pm0.20$ &\hfil $4.63\pm0.15$&\\
            \hfil $A/R_{\odot}$&\hfil $2.03\pm0.02$ &\hfil $2.16\pm0.01$&\\
            \hfil $R_1/R_{\odot}$&\hfil $1.19\pm0.02$ &\hfil $1.27\pm0.02$&\\
            \hfil $R_2/R_{\odot}$&\hfil $0.46\pm0.02$ &\hfil $0.49\pm0.02$&\\
            
            \hline
           
    \end{tabular}
    
    \end{table}

 \subsection{Orbital stability of W1908 and A2044}
 
 The knowledge of the mass of the primary is essential for determining orbital stability. As direct measurement of the mass of the primary is not possible, we have previously employed the mean of an infrared ($J-H$) colour calibration and a distance-based absolute magnitude calibration. The colour calibration uses catalogued $J$ and $H$ band photometry along with the April 2022 update calibrations from \citet{2013ApJS..208....9P} for low mass main sequence stars. The absolute magnitude method uses the mid eclipse apparent magnitude (which represents the apparent magnitude of the primary) corrected for extinction and GAIA EDR 3 \citep{2022A&A...658A..91A} distances along with same calibration tables. Detailed methodology can be found in \citep{2023PASP..135g4202W}. We determine the mass of the primary of W1908 to be $(1.06\pm0.02)\rm M_{\odot}$ and for A2044 to be $(1.04\pm0.02)\rm M_{\odot}$. The distance based estimate of the mass yields the largest error and this was adopted and propagated when estimating errors of other parameters. Given the similarity in the mass of the primaries, it is not surprising that the SED estimates for their effective temperatures were the same. The mass ratio yields the mass of the secondary and Kepler's 3rd law allows determination of the separation ($A$). The light curve solution provides fractional radii of the components ($r_1, r_2$) in different orientations and we use the geometric mean of these to estimate the absolute radii as $(R_1, R_2)$ = $A\cdot(r_1, r_2)$. The absolute parameters are summarised in Table 5.

 As described by \citep{2021AJ....162...13L}, standard black body calibrations along with distance and luminosity can be used to estimate absolute parameters. Using this approach with our determined values of the absolute magnitude of the primaries and estimated temperatures of $6000K$ we estimate the luminosity of the primary component ($L_1$) of W1908 as $0.98L_{\odot}$ and its radius ($R_1$) as $0.92R_{\odot}$ and its mass ($M_1$) as $0.94M_{\odot}$. Similarly, for A2044 we estimate $L_1, R_1$ and $M_1$ as $1.17L_{\odot}$, $1.00R_{\odot}$  and $1.00M_{\odot}$. We prefer to use estimates based on the geometric light curve solution principally due to the dependency of the black body estimates on the estimated effective temperature. As noted by \citet{2023arXiv230811906W} a 200K variation in the assigned value of $T_1$ can lead to a greater than 10\% change in the estimated value of $M_1$ for low mass stars. The corresponding change in the instability mass ratio can exceed 15\%. Additionally, a number of steps are required to determine luminosity, radius and then mass; each associated with its own error which would require propagation leading to a larger overall error in the estimate. Lastly, the black body approximation is based on a spherical configuration, it is well known that contact binary star components are not spherical and considerably distorted by the Roche geometry such that the mean radius of both the primary and secondary are considerably larger than their main sequence counterparts \citep{2022JApA...43...94W}.
 
 At $\rm [Fe/H] = 0$ the instability mass ratio range for W1908 is $q_{\rm inst} = 0.093\pm0.004 - 0.108\pm0.004$ and for A2044 it is $q_{\rm inst} = 0.096\pm0.004 - 0.112\pm0.004$. As expected the values are similar to the simplified relationships described in \citet{2021MNRAS.501..229W} (0.091 - 0.106 for W1908 and 0.095 - 0.110 for A2044). The modelled mass ratios for both systems are within the errors for the instability mass ratio range and both would be considered unstable and potential merger candidates. Unfortunately no spectroscopic metallicity estimates are available for either system. The GAIA EDR3 does report photometric metallicity for W1908 as $\rm [Fe/H] = -1.04$ and for A2044 as $\rm [Fe/H] = -0.38$. If we use the nearest metallicity corrected values for the gyration radii for each system $\rm [Fe/H] = -1.0$ for W1908 and $\rm [Fe/H] = -0.375$ for A2044, then the instability mass ratios reduce to $0.068 - 0.076$ for W1908 and $0.084 - 0.097$ for A2044. Both systems would now be classified as stable with modelled mass ratios well above the instability range. However as discussed by \citet{2023A&A...674A..27A} the photometric metallicity estimates provided by GAIA EDR3 exhibit systematic errors. As a result, they should not be used for quantitative analysis without a suitable calibration. \citet{2023A&A...674A..27A} provide a calibration trained on a large sample of LAMOST spectra and we use this to adjust the metallicities to $\rm [Fe/H] = -0.7$ for W1908 and $\rm [Fe/H] = -0.03$ for A2044. Based on the revised metallicity values, the instability range for W1908 is $0.074 - 0.084$, still significantly below the modelled mass ratio, and the system is considered stable. The revised metallicity of A2044 is near zero such that the instability mass ratio is as initially determined and the system would be considered unstable and a merger candidate. This exercise demonstrates that two very similar systems in many respects such as the mass of the primary and the geometric parameters can have very different orbital stability indicators relating to internal stellar structure. Metallicity, we suggest, must be taken into consideration when classifying any contact binary system as potentially unstable.

 \section{Summary and Conclusion}
Identification of potential red nova candidates has been receiving much attention recently \citep{2021MNRAS.501..229W, 2021MNRAS.502.2879G, 2022MNRAS.512.1244C, 2023MNRAS.519.5760L}. It is well understood that merger will only occur in systems with low mass ratios although \citet{2021MNRAS.501..229W} showed that systems where the primary is less than one Solar mass may become unstable at mass ratios higher than previously reported. It has long been recognised that orbital stability is dependent on the gyration radii of the components. As the secondary is very small, being fully convective, it is reasonable to assume a fixed value for its gyration radius \citep{2007MNRAS.377.1635A, 2009MNRAS.394..501A}. The effect of the gyration radius of the primary on orbital stability is only now beginning to receive attention and early indications are that it is significant \citep{2021MNRAS.501..229W}. The gyration radius enters the definition of moment of inertia so it would be affected by the internal structure and composition of a star. The effects of metallicity on orbital stability have received scant attention previously.

It is well known that different compositions of a star will alter its gyration radius \citep{2004A&A...424..919C, 2019A&A...628A..29C}. In this study we model the gyration radii of low mass stars incorporating rotation and tidal distortion effects at a range of metallicities from $-1.25\leq \rm [Fe/H] \leq 0.5$. We then apply these to orbital stability equations to show that metal content indeed has a significant effect on the instability parameters of contact binary stars. The main observation is that the instability mass ratio decreases with decreasing metallicity, however this does not hold true for stars with masses greater than $1\rm M_{\odot}$ at lower metallicity levels. The trend is similar to that seen for stars heavier than $1.5\rm M_{\odot}$ at Solar metallicity levels. It is likely that the change is due to transition to the CNO energy cycle which we show is likely in stars of mass $\approx 1.1\rm M_{\odot}$ at metallicity below $\rm [Fe/H] = -1$. 

Low mass ratio contact binary systems are uncommon. \citet{2021ApJS..254...10L} catalogued light curve and radial velocity solutions of over 680 contact binary systems with less than 150 having a mass ratio below 0.25. Review of the LAMOST spectra catalogued by \citet{2020RAA....20..163Q} suggests that over 70\% of contact binary systems in the solar neighborhood (most likely to be available for long term monitoring) have metallicities less than zero. Estimates suggest that Galactic frequency of observable red nova events should be about once every 10 years, however, there still remains only one confirmed (in retrospect) contact binary merger event, that of V1309 Sco. It is possible that such events are in fact rare in the Solar neighborhood where contact binaries have lower metal content and are likely to be in a more stable configuration. Although young open clusters may provide an avenue for the detection of high metal content contact binaries, their inherent low mass would make them difficult to detect without time consuming observations requiring relatively large instruments which is likely prohibitive. The survey nature of the Extremely Large Telescope observing programme, however, may provide a tremendous opportunity to observe faint potentially unstable contact binary systems.

\section*{Acknowledgements}

\noindent This research has made use of the SIMBAD database, operated at CDS, Strasbourg, France.\\

\noindent N.R. Landin acknowledges financial support from Brazilian agencies FAPEMIG, CNPq and CAPES.\\

\noindent This publication makes use of VOSA, developed under the Spanish Virtual Observatory (https://svo.cab.inta-csic.es) project funded by MCIN/AEI/10.13039/501100011033/ through grant PID2020-112949GB-I00. VOSA has been partially updated by using funding from the European Union's Horizon 2020 Research and Innovation Programme, under Grant Agreement number 776403 (EXOPLANETS-A).\\

\noindent B. Arbutina acknowledges the funding provided by the Ministry of Science, Technological Development and Innovation of the Republic of Serbia through the contract  451-03-47/2023-01/200104.\\

\noindent  O. Vince and P. Kosti\'c acknowledge support by the Astronomical station Vidojevica, funding from the Ministry of Science, Technological Development and Innovation of the Republic of Serbia (contract No. 451-03-47/2023-01/200002), by the EC through project BELISSIMA (call FP7-REGPOT-2010-5, No. 265772).\\

\noindent During work on this paper, G. Djurašević and J. Petrović were financially supported by the Ministry of Science, Technological Development and Innovation of the Republic of Serbia through contract 451-03-47/2023-01/200002\\

\section*{Data Availability}
The data underlying this article will be shared on reasonable request to the corresponding author.




\bibliographystyle{mnras}
\bibliography{SSW-bibtex} 





\bsp	
\label{lastpage}
\end{document}